\documentclass[twocolumn,showpacs,preprintnumbers,amsmath,amssymb]{revtex4}
\usepackage{graphicx}
\begin{document}
\preprint{IMAFF-RCA-03-03}
\title{Axion Phantom Energy}
\author{Pedro F. Gonz\'{a}lez-D\'{\i}az}
\affiliation{Centro de F\'{\i}sica ``Miguel A. Catal\'{a}n'', Instituto de
Matem\'{a}ticas y F\'{\i}sica Fundamental,\\ Consejo Superior de
Investigaciones Cient\'{\i}ficas, Serrano 121, 28006 Madrid (SPAIN).}
\date{\today}
\begin{abstract}
The existence of phantom energy in a universe which evolves to
eventually show a big rip doomsday is a possibility which is not
excluded by present observational constraints. In this letter it
is argued that the field theory associated with a simple
quintessence model is compatible with a field definition which is
interpretable in terms of a rank-three axionic tensor field,
whenever we consider a perfect-fluid equation of state that
corresponds to the phantom energy regime. Explicit expressions for
the axionic field and its potential, both in terms of an imaginary
scalar field, are derived which show that these quantities both
diverge at the big rip, and that the onset of phantom-energy
dominance must take place just at present.
\end{abstract}


\maketitle

\section{Introduction}

Cosmology has become the current commonplace where the greatest
problems of physics are being concentrated or clearest manifest
themselves. Along with the nature of dark matter, the origin of
supermassive black holes and several unexplained huge amounts of
released astrophysical energy are only some illustrative examples.
However, perhaps the biggest problem of all of physics be the
so-called dark energy problem [1] by which it is known that nearly
the seventy percent of the total energy in the universe is in the
form of an unobserved vacuum energy which is responsible for the
present accelerating expansion of the universe [2,3] and whose
nature remains being a mystery today [4]. There has been a rather
huge influx of papers in recent years trying to shed some light
onto the existence and the kind of the possible stuff that may
make up dark energy [5].

Four main candidates to represent dark energy have been hitherto
suggested: A positive cosmological constant [6], the quintessence
fields (which may [7] or may not [8] be tracked), some
generalizations of the Chaplygin gas [9], and the so-called
tachyon model of Padmanabhan {\it et al.} [10]. Even though such
models could be all accommodated to present observational
constraints, these models are posing new or traditional problems
in such a way that none of them become completely satisfactory. In
particular, cosmological data from current observations do not
exclude [11] but may be suggesting [12] values of the parameter
$\omega$ in the perfect-fluid equation of state $p=\omega\rho$ of
the most popular quintessence models which are smaller than -1. If
this were the case, then dark energy would become what is now
named phantom energy [13], an entity violating the dominant energy
condition and thereby allowing the natural occurrence of wormholes
and ringholes and even their corresponding time machines in the
universe [14].

Caldwell, Kamionkowski and Weinberg have recently noticed [15]
that in the framework of quintessence models phantom energy may
lead to a doomsday for the universe - which would take place at a
big rip singularity - once clusters, galaxies, stars, planets and,
ultimately, nucleons and leptons in it are all ripped apart. Even
though such a big rip does not take place in some phantom-energy
models that use Chaplygin-like equations of state [16], it seems
to be the simplest and most natural possibility stemming from
phantom energy. On the other hand, the big rip cosmic scenario
interestingly adds an extra qualitative feature to the known set
of cosmological models as it introduces a curvature singularity,
other than that at the big bang, at a finite, nonzero value of the
cosmic time. However, in spite of the feature that the blowing up
of the scale factor appears to be unavoidable in models with
equation of state $p=\omega\rho$ and $\omega<-1$, a complete
account of the nature and properties of the field theory
associated with such models has not been done yet. This paper aims
at investigating the characteristics of the massless scalar field
which allow emergence of a big rip singularity in the simplest of
such phantom energy models, assuming a perfect-fluid equation of
state. It will be shown that the stuff making up phantom energy
can be interpreted to be a vacuum sea of cosmic axions which can
be described in terms of the kind of rank-three tensor field
strengths predicted in supergravity and strings theories.

This paper can be outlined as follows. In Sec. II we argue in
favor of the idea that superlight axions are the source of cosmic
phantom energy. A simple cosmological model accounting for a big
rip singularity in the case that vacuum is filled with phantom
energy is discussed in Sec. III. Sec. IV contains the solution of
the phantom field theory within the simple cosmological model of
Sec. III. We check that both the scalar field and its potential
also have a singularity at the big rip. The results are summarized
in Sec. V.

\section{Axions as the source of phantom energy}

The definition of the massless scalar field $\phi$ which is
assumed to make up dark energy is usually taken to be the
conventional simplest one; that is, in terms of pressure $p$ and
energy density $\rho$,
\begin{equation}
\rho=\frac{1}{2}\dot{\phi}^2 +V(\phi) ,\;\;\;
p=\frac{1}{2}\dot{\phi}^2 -V(\phi) ,
\end{equation}
where $V(\phi)$ is the field potential. From the equation of state
$p=\omega\rho$ and Eq. (1) it immediately follows that
\begin{equation}
\rho=\frac{\dot{\phi}^2}{1+\omega} .
\end{equation}
Thus, if the weak energy condition $\rho\geq 0$ is taken to be
always satisfied [17], the requirement $p+\rho<0$, $\omega<-1$
from phantom energy in this kind of models necessarily implies
that the field $\phi$ ought to be pure imaginary; that is to say,
if the vacuum energy density for a phantom vacuum as referred to
any timelike observer has to be positive, then the massless scalar
field making up the phantom stuff should be pure imaginary. I will
argue in what follows that, in the classical framework, such a
massless, pure imaginary scalar field actually represents an axion
describable as a rank-three antisymmetric tensor field,
considering after the associated cosmic theory for such a field.
In fact, the Lorentzian action that couples such an axion field to
gravity can generally be written as
\begin{equation}
S=\int d^4 x\sqrt{-g}\left(\frac{R}{16\pi G}-{\it A}^2+L_m\right)
,
\end{equation}
where $L_m$ is the Lagrangian for observable matter and ${\it
A}^2\equiv{\it A}_{\mu\nu\alpha}{\it A}^{\mu\nu\alpha}$ if we
choose the axion to be given by a three-form ${\it A}=dB$ field
strength so that $d{\it A}=0$, that is as a rank-three
antisymmetric tensor field strength of a type arising in
supergravity-theory motivated quantum-gravity solutions [18]. The
equations of motion derived from action (3) are
\begin{equation}
R_{\mu\nu}-\frac{1}{2}g_{\mu\nu}R=16\pi G\left(3{\it A}_{\mu\nu}^2
-\frac{1}{2}g_{\mu\nu}{\it A}^2 + T_{\mu\nu}^{(m)}\right)
\end{equation}
\begin{equation}
d*{\it A}=0 ,
\end{equation}
in which ${\it A}_{\mu\nu}^2={\it A}_{\mu\alpha\beta}{\it
A}_{\nu}^{\alpha\beta}$, the asterisk denotes Hodge dual, and
$T_{\mu\nu}^{(m)}$ is the momentum-energy tensor for ordinary
observable matter. One now can check that any explicit solution to
these equations of motion subject to the usual
Friedmann-Robertson-Walker (FRW) symmetry for the metric and the
corresponding spherically symmetric ansatz for the axion
antisymmetric tensor ${\it A}=f(r)\epsilon$ (with $\epsilon$ the
volume form which, when integrated over a surface of constant
radius, yields the area of the unit three-sphere ${\it
A}=2\pi^2/\Gamma(2)$), if we set e.g. $L_m=0$ and $f(r)={\rm
Const.}$, is the same as the solution obtained from the equations
of motion derived from the FRW action integral describing coupling
of a real massless scalar field $\phi$ to gravity where an extra
boundary term accounting for the axion properties,
$-a^3\dot{\phi}\phi|_0^T$ ($a$ being the scale factor), is added;
i.e. in the flat geometry case
\begin{equation}
S'=\frac{1}{16\pi G}\int_0^T dt a^3\left[-\frac{\dot{a}^2}{N}+8\pi
G\frac{\dot{\phi}^2}{N}\right]-
\left.\frac{a^3\dot{\phi}\phi}{N}\right|_0^T ,
\end{equation}
where $N$ is the lapse function. The key point then is that the
solution to the above equations of motion is even also preserved
when we omit such an extra boundary term in the action for the
scalar field $\phi$, provided this scalar field is simultaneously
rotated to the pure imaginary axis, $\phi\rightarrow i\Phi$. If
the scalar field is equipped with a potential $V(\phi)$, then an
extra term $-\int_0^T dtNA^3 V(\phi)$ should be added to the
action. In that case, the axionic action can be obtained by simply
rotating $\phi\rightarrow i\Phi$ in an action containing the above
potential term, without any extra boundary term.

Such an property, which was first noticed in Euclideanized
solutions such as wormholes and other instantons describing
nucleation of baby universes [19], is actually independent of the
metric signature and, therefore, applies also to our Lorentzian
cosmological context. Thus, one can interpret that the stuff that
makes up phantom energy can be regarded to be a rank-three
antisymmetric tensor axionic field. It is usually thought that
axions can explain the absence of an electrical dipole moment for
the neutron and thereby solve the so-called strong CP problem
[20]. The axions are chargeless and spinless particles with very
tiny mass which interact with ordinary matter only very weakly.
Such particles are believed to have been abundantly produced in
the big bang. It is worth noticing that whereas relic axions are
an excellent candidate for the dark matter in the universe [21],
their vacuum quantum background could make cosmic phantom energy.

The coincidence and fine tuning problems could be thought to
become exacerbated in the present scenario where one sets $\omega$
constant and $<-1$. However, in dark energy models such as the
generalized Chaplygin gas [9] and tachyon models [10], dark energy
and dark matter are described as separate limiting cases from an
existing unique field. Partly inspired by these models one could
naively assume the existence of a unique axion field which, when
excited, would make dark matter, and when at its vacuum ground
state would be the source of phantom energy. Coincidence time
could then be interpreted as the time when both the vacuum and the
excited states are approximately equally populated. Of course,
this would not solve the coincidence and fine tuning problems but
provided some explanation to these problems and to the prediction
of generating such strangely small amounts of homogeneously
distributed axions. It remains nevertheless an intriguing
possibility that such a small value of the cosmological constant
be supplied by the potential energy density of an "ultrainvisible"
axion field which can be dubbed quintessence axion [22].

\section{The cosmological model}

We shall consider in what follows a simple model where the
massless scalar field $\phi$ is a quintessential field, equipped
with a potential $V(\phi)$, which is minimally coupled to
Hilbert-Einstein gravity. The action for this model is
\begin{equation}
S_c= \int d^4 x\sqrt{-g}\left(\frac{R}{16\pi G}
+\frac{1}{2}\partial^i\phi\partial_i\phi-V(\phi)+L_m\right) ,
\end{equation}
where $L_m$ again is the Lagrangian for observable matter fields.
If we again restrict ourselves to the case where (i) $L_m=0$, (ii)
both the scalar field and Hilbert-Einstein gravity satisfy the
Friedmann-Robertson-Walker symmetry, and (iii) we assume a perfect
fluid equation of state, then by integrating the conservation law
for cosmic energy, $d\rho=-3(p+\rho)da/a$, it turns out that the
energy density for the quintessence scalar field will be given by
\begin{equation}
\rho=Ra^{-3(1+\omega)} ,
\end{equation}
in which $R$ is an integration constant. From the Friedmann
equation derived from action (7) subject to the FRW symmetry and
$L_m=0$,
\begin{equation}
\left(\frac{\dot{a}}{a}\right)^2=Aa^{-3(1+\omega)}, \;\;\;
A=\frac{8\pi GR}{3} ,
\end{equation}
we can obtain the solution
\begin{equation}
a(t)=\left[a_0^{3(1+\omega)/2}+
\frac{3(1+\omega)\sqrt{A}}{2}(t-t_0)\right]^{\frac{2}{3(1+\omega)}}
,
\end{equation}
where $a_0$ and $t_0$ are the initial radius and time,
respectively. Note that for $\omega >-1$ this solution describes
an accelerating universe whose scale factor increases towards
infinity as $t\rightarrow\infty$. We are here most interested in
the observationally not excluded yet case where $\omega <-1$ which
corresponds to the so-called phantom dark energy for which the
dominant energy condition is generally violated, i.e. [15]
\begin{equation}
p+\rho <0 ,
\end{equation}
even though the energy density is surprisingly ever increasing.
Notice furthermore that in this case the scale factor blows up at
a finite time,
\begin{equation}
t_* =t_0+\frac{2}{3(|\omega|-1)a_0^{3(|\omega|-1)/2}} ,
\end{equation}
giving rise to a "big rip" [15]. One would expect that in the
vecinity of time $t_*$ the size of all local geometrical objects
increased in a rather dramatic way. We see that the larger $a_0$
and $|\omega|$ the nearer the doomsday, and hence the time
expected to have macroscopically increased geometrical objects in
our universe.

\section{The field theory}

Now we shall consider the field theory which is associated with
phantom energy, starting with the general formalism for any value
of $\omega$. From the equation of motion for the scalar field
$\phi$,
\begin{equation}
\ddot{\phi}+3H\dot{\phi}=-\frac{dV(\phi)}{d\phi} ,
\end{equation}
where $H\equiv\dot{a}/a$ and $V(\phi)$ again is the potential for
the field $\phi$, and the expression for kinetic energy of the
field [Eq. (2)], we can obtain a relation between $dV/d\phi$ and
the scale factor which can be expressed in the form
\begin{equation}
\frac{3}{2}\sqrt{A(1+\omega)}(1-\omega)a^{-3(1+\omega)}
=-\frac{dV}{d\phi} .
\end{equation}
On the other hand, integrating the equation of motion for the
field $\phi$ we can derive the dependence of this field with time
$t$.
\begin{equation}
\phi(t) = \frac{2}{3\sqrt{1+\omega}\ell_p}
\ln\left[\frac{a_0^{3(1+\omega)/2} +\frac{3(1+
\omega)\sqrt{A}}{2}(t-t_0)}{B_0}\right] ,
\end{equation}
where $\ell_p=\sqrt{R/A}$ is the Planck length and $B_0$ is an
integration constant.

\begin{figure}
\includegraphics[width=.9\columnwidth]{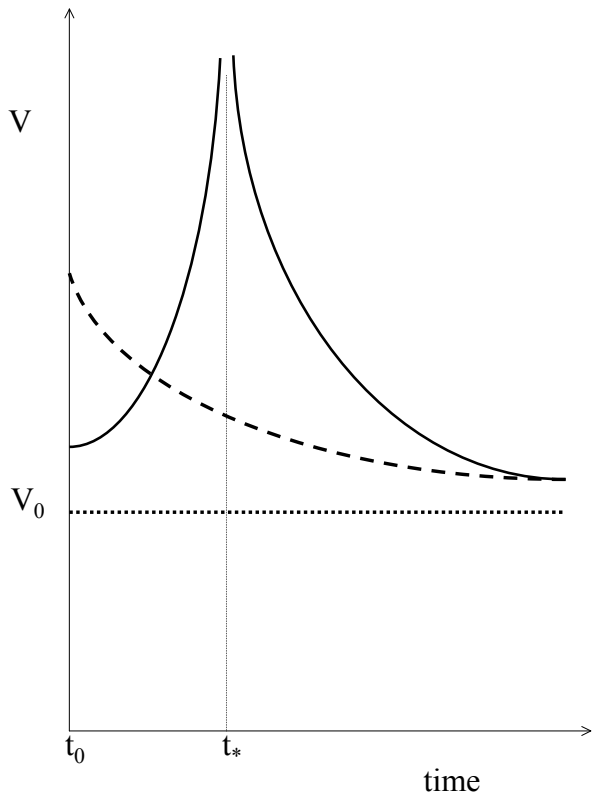}
\caption{\label{fig:epsart} The potential $V$ for imaginary (solid
lines) and real (dashed line) scalar field. The scales for the
potential and time are arbitrary. $t_*$ is the time at which the
big-rip singularity would occur in the future. It is worth
noticing that the singularity in the spacetime curvature at $t_*$
coincides with a singularity in the imaginary field $\Phi$ and the
potential $V(\Phi)$. In constructing this figure it has been
assumed that the initial scale factor and the parameter entering
the respective equations of state are such that $a_0 >>
[(1-\omega_d)/(1+|\omega_p|)]^{1/[3(\omega_d+|\omega_p|)]}$, where
$\omega_d>-1$ and $\omega_p<-1$, with $a_0$ sufficiently large.}
\end{figure}

From Eqs. (10) and (15) an expression for the scale factor in
terms of the scalar field can now be derived. It is:
\begin{equation}
a(\phi)=
B_0^{\frac{2}{3(1+\omega)}}\exp\left(\frac{\ell_p\phi}{\sqrt{1+\omega}}\right).
\end{equation}
Using Eqs. (14) and (16) we finally obtain the expression of the
field potential for any $\omega<-1/3$ as a function of the field
$\phi$
\begin{equation}
V(\phi)= V_0 + \frac{1}{2}(1-
\omega)\sqrt{R}B_0^{-2}\exp\left(-3\sqrt{1+\omega}\ell_p\phi\right)
,
\end{equation}
in which $V_0$ is a constant. Potentials with this form naturally
arise in supergravity models [23] and have been used in a variety
of context, ranging from accelerating expansion models [24] to
cosmological scaling solutions [25].

Together with the expression for the field $\phi$ in terms of time
$t$, this expression solves the field-theory problem for any value
of $\omega<-1/3$. According to our discussion above, for axionic
phantom energy we have to perform the continuation
$\phi\rightarrow i\Phi$ in addition to taking $\omega<-1$. In such
a case, the solution of the field theory problem would read
\[\Phi(t) =\]
\begin{equation}
-\frac{2}{3\sqrt{|\omega|-1}\ell_p}
\ln\left[\frac{a_0^{-3(|\omega|-1)/2}
-\frac{3(|\omega|-1)\sqrt{A}}{2}(t-t_0)}{B_0}\right]
\end{equation}
\begin{equation}
V(\Phi)= V_0 + \frac{1}{2}(1+
|\omega|)\sqrt{R}B_0^{-2}\exp\left(3\sqrt{|\omega|-1}\ell_p\Phi\right)
.
\end{equation}

As to the physical motivation for the exponential axion potential
(19), I will briefly comment on the influence that the
matter-field sources may have on the stability of the present
axion model of phantom energy. If, corresponding to the physically
interesting case that $V_m\neq 0$, we would let the influence of
matter sources on the expansion factor immediately before of
phantom domination to have the form of a perturbation, and analyze
the resulting model via phase space [26], it can be seen that if
we choose $V_0=0$ then potential (19) and the solution given by
Eq. (10) (with $\omega<-1$) and Eq. (18), that is for
\[t=t_0 +\frac{2\left[1-
\left(\frac{a_0}{a}\right)^{3(|\omega|-
1)/2}\right]}{3a_0^{3(|\omega|-1)/2}\left(|\omega|-
1\right)\sqrt{A}} ,\] correspond (see e.g. Ref. [26]) to a
dynamical attractor for phantom energy domination and catastrophic
big rip with $\omega<-1$ and
\[\Gamma=\frac{V(\Phi)' V(\Phi)''}{\left[V(\Phi)'\right]^2}=1 ,\]
at a critical point defined by
\[8\pi G\lambda_c =-\frac{V(\Phi)'}{V(\Phi)}=
-3\sqrt{|\omega|-1}\ell_p ,\] for which $\omega=-1-\lambda_c^2/3$
(See also phantom models with Born-Infeld type Lagrangians [27].)

Note that the potential $V(\Phi)$ takes on a value
\begin{equation}
V=V_0+\frac{1}{2}(1+|\omega|)\sqrt{R}a_0^{3(|\omega|-1)}
\end{equation}
at the initial time $t=t_0$ to steadily increase thereafter up to
infinity when $t=t_*$, at the big rip. From that moment on the
potential would continuously decrease, tending to reach its
constant minimum value $V_0$ as $t\rightarrow\infty$ (see solid
curves in Fig. 1). That behaviour is in sharp contrast with that
of potential $V(\phi)$ for dark energy with $\omega>-1$, which
starts with a value
\[V=V_0+\frac{1}{2}(1-\omega)\sqrt{R}a_0^{-3(1+\omega)} \]
to monotonously decrease down to the value $V_0$ as
$t\rightarrow\infty$, such as it is also shown by the dashed curve
in Fig. 1. We have solved in this way the field theory which is
associated with dark and phantom energy if the latter is
interpreted as originating from the existence of an axionic field
which is expressible as a massless, purely imaginary scalar field.
The explicit emergence of axions in the cosmological context of
phantom dark energy appears to be just another example where such
rather elusive particles are invoked as candidates for inexorably
needed cosmic constituents which are themselves defined to be
unobservable. Actually, apart of being a key ingredient to solve
the strong-CP problem [20] or to represent quantum-gravity
topology changes [19], axions have already been widely used to
solve a variety of astrophysical and cosmological shortcomings
[28].

It appears of some interest to briefly comment next on a
potentially fruitful consequence from the connection of axions
with our cosmological model. If we are actually living in a
universe with phantom energy where there will occur a big rip
singularity which can never be circumvented by causally-violating
connections to the future, then the contracting branch of solution
(9) can be physically disregarded as the big-rip singularity is a
true curvature singularity. In such a case, only the branch of
potential $V(\Phi)$ before the singularity can have physical
significance, and hence the global minimum of that potential would
become placed at $t=t_0$ with the value given by Eq. (20). It
corresponds to a value of the field
\begin{equation}
\Phi_{{\rm m}}= \frac{2}{3\ell_p\sqrt{|\omega|-1}} \ln\left(B_0
a_0^{3(|\omega|-1)/2}\right) .
\end{equation}
Interpreting potential (19) as that axion potential resulting from
the QCD nonperturbative effects, we can take [29] $\Phi_{{\rm
m}}=f_A \theta_{{\rm eff}}$, with $f_A$ a constant and
$\theta_{{\rm eff}}$ the effective $\theta$ parameter resulting
after the diagonalization of the quark masses in the QCD
Lagrangian. Setting then $f_A= 2/[3\ell_p\sqrt{|\omega|-1}]$,
$\theta_{{\rm eff}}=\ln\left(B_0 a_0^{3(|\omega|-1)/2}\right)$ and
$B_0=1$ it follows that, since $\theta_{{\rm eff}} < 10^{-9}$
[29], the initial radius of the phantom energy universe $a_0$
should be very close to unity, that is the onset of such a phantom
energy regime, if it ever at all occurred, must always be placed
at nearly just the present epoch. Actually, if $\theta_{{\rm
eff}}$ is cancelled to completely solve the strong CP problem
[29], then the choice $B_0=1$ implied that $a_0$ would exactly
satisfy $a_0=1$. In such a case, the observational setting of the
value of the state equation parameter $\omega$ would also set the
value of $a_0$, and hence of time $t_*$.

A potential problem with the proposal in this paper could be that
whereas current axion theories have a cut off on the scale of
inflationary energy, typically at around the GUT characteristic
energy of $10^{16}$ GeV [29], in order to avoid phantom energy
decaying into gravitons, a cut off of $< 100$ MeV should be
introduced in the effective phantom theory [30]. Not with
standing, while the former cut off appears relevant for dark
matter axions which can initially be highly excited, the axion
field proposed in this paper as the source of phantom energy can
never be excited outside the vacuum. On the other hand, if we keep
the current values for color anomaly of the Peccei-Quinn symmetry,
the pion and quark masses and the pion decay constant, then the
above Peccei-Quinn symmetry-breaking scale, $f_A\sim
10^{19}(|\omega|-1)^{-1}$ GeV, would imply a mass of the phantom
axions
\[m_A \sim 0.62\; {\rm eV} \frac{10^7\; {\rm GeV}}{f_A} \sim
10^{-12}\sqrt{|\omega|-1}\; {\rm eV} ,\] i.e. just at the extremal
minimum of the allowed values for the axion mass, and clearly well
below the cut off required by Carroll et al. [30] for phantom
energy to be stable.

\section{Summary}

Imposing the weak energy condition, the negative kinetic energy
for a phantom field is interpreted as being originated from
super-light axions. We have built up a simple cosmological model
encompassing constant equations of state for both $\omega >-1$ and
$\omega<-1$. In the latter case the universe evolves toward a
singularity at finite time. The field theory is then solved for
all cases, deriving an increasing exponential potential for the
phantom field which also diverges at the singularity and shows a
dynamical attractor also when matter fields are present that
inexorably leads to a catastrophic big rip.

\acknowledgements

\noindent The author thanks Carmen L. Sig\"{u}enza for useful
discussions. This work was supported by DGICYT under Research
Project BMF2002-03758.

\end{document}